\documentclass[twocolumn,aps,pra,tightenlines,amsmath,amssymb]{revtex4-1}
\usepackage{graphicx}
\usepackage{amssymb}
\usepackage{dcolumn}
\usepackage{amsmath}
\usepackage{bm}
\usepackage{colordvi}

\begin{document}

\title{Weak Measurements Beyond the Aharonov-Albert-Vaidman Formalism}
\author{Shengjun Wu and Yang Li}

\affiliation{Hefei National Laboratory for Physical Sciences at Microscale and
Department of Modern Physics,
University of Science and Technology of China,
Hefei, Anhui 230026, China}

\date{\today}

\begin{abstract}
We extend the idea of weak measurements to the general case, provide a complete treatment and obtain results for both the regime when the pre-selected and post-selected states (PPS) are almost orthogonal and the regime when they are exactly orthogonal. We surprisingly find that for a fixed interaction strength, there may exist a maximum signal amplification and a corresponding optimum overlap of PPS to achieve it.  For weak measurements in the orthogonal regime, we find interesting quantities that play the same role that weak values play in the non-orthogonal regime.
\end{abstract}


\maketitle

\section{Introduction}  
Weak measurements, which were introduced by Aharonov, Albert and Vaidman (AAV) in 1988 \cite{AAV} and implemented successfully in the experiments reported in Refs. \cite{Ritchie,Pryde05,HK2008}, have challenged the view that the value of an observable in quantum mechanics has physical reality only if it is actually measured. The idea of weak measurement is very useful as it assists us in understanding many counterintuitive quantum phenomena, resolving paradoxes that arise in quantum mechanics such as Hardy's paradoxes \cite{ABPRT02}, and addressing questions of fundamental importance \cite{Wiseman07,Mir07,WJ08,GABLOWP09,Johansen}.
It was realized recently that weak measurements are also very useful for high-precision measurements, they are used by Hosten et al. \cite{QSHE} to observe a tiny spin Hall effect in light (see a nice perspective in \cite{Resch}),
and by Dixon et al. \cite{Dixon, Starling} to detect very small transverse beam deflections.
A weak-measurement scheme was also recently proposed by Brunner et al. \cite{BS} to measure small longitudinal phase shifts.

Weak measurement of an observable $\mathbf{A}$ typically includes  pre-selection of an initial state for the system, weak interaction between the system and a measuring device, and post-selection of a final state for the system.
The interaction, described by the Hamiltonian $\mathcal{H}=g(t-t_0)\mathbf{A} \otimes p $, associates a certain property of the system observable $\mathbf{A}$
to the shift of a certain pointer position $q$ of the measuring device.
Here $q$ and $p$ are a pair of conjugate variables of the device, satisfying $[q,p]=i$ (with $\hbar =1$ throughout this article).
$g(t-t_0)$ is a real function with $\int{g(t-t_0)dt}=g$, and it is frequently given in the simple form $g(t-t_0)=g\delta(t-t_0)$.
For an ideal measurement, the initial state of the device is well localized in the $q$ space so the pointer position $q$ will be shifted by an amount determined by one of the eigenvalues of $\mathbf{A}$ with a certain probability.
Weak measurement takes the opposite approach, we either choose an initial pointer state that is spread out greatly over the $q$ space (with a small standard deviation $\Delta p$ in the p space) or choose a very small $g$, such that \cite{AAV}
\begin{equation}
g \Delta p   \ll \frac{|\langle \psi_f  | \psi_i \rangle|}{|\langle \psi_f |\mathbf{A}^n | \psi_i \rangle|^{1/n}} \text{ for } n=1,2,\cdots \label{AAVcondition}
\end{equation}
where $\left| \psi_i \right\rangle$ is the pre-selected state and $\left| \psi_f \right\rangle$ is the post-selected state of the system according to AAV's original formalism.
A subsequent ideal projection measurement is preformed, and we only keep the subset of data that corresponds to the post-selection.
Conditional on the pre- and post-selection, the average shifts of $q$ and $p$ of the measuring device in a weak measurement are given by \cite{AAV, Jozsa}
\begin{eqnarray}
\delta q &=& \left\langle q \right\rangle ' - \left\langle q \right\rangle =  g \Re \mathbf{A}_w + g\Im \mathbf{A}_w \cdot \langle \lbrace p,q \rbrace \rangle  \label{deltaq} \\
\delta p &=& \left\langle p \right\rangle ' -\left\langle p \right\rangle  =  2 g \Im \mathbf{A}_w \cdot \text{var} p
\label{deltap}
\end{eqnarray}
where $\mathbf{A}_w$ is the weak value defined by
\begin{equation}\label{weak value}
\mathbf{A}_w = \frac{\langle \psi_f | \mathbf{A} |\psi_i \rangle}{\langle \psi_f|\psi_i\rangle} .
\end{equation}
Here and in the following, $\left\langle \hat{o} \right\rangle$ ($\text{var} \hat{o}$) denotes the expectation value (variance) of an observable $\hat{o}$ of the device in its initial state $\phi$, and $\left\langle \hat{o} \right\rangle '$ ($\text{var}' \hat{o}$) with a prime denotes the corresponding value in the final state of the device after the interaction and post-selection.  $\Re Z$ and $\Im Z$ denote the real part and imaginary part of a complex number $Z$ respectively.

The high-precision experiments via weak measurement have been performed with a post-selected state nearly orthogonal to the pre-selected state, as the weak value $\mathbf{A}_w$ could become very large when the pre- and post-selected states (PPS) are close to being orthogonal. However, it should be pointed out that when the PPS are too close to being orthogonal,
condition (\ref{AAVcondition}) is no longer satisfied and the results in (\ref{deltaq}) and (\ref{deltap}) start to fail.  There are some specific discussions about the phenomena in this regime \cite{Duck89,SDWJH,Geszti}, but much needs to be understood and a complete treatment is lacking.

In this article we shall study weak measurements in the most general case, provide theoretical results to cover the regime that is potentially relevant in certain high-precision measurements.
Our study also reveals two surprising results. First, for a fixed interaction strength, there may exist a maximum measurement outcome and an optimum overlap for the PPS to achieve it. Second, even when the PPS are exactly orthogonal, there exist interesting quantities that play a role similar to that of the weak values.
The rest of the article is arranged as follows. We first introduce the weak-interaction condition and then set up a framework for discussing weak measurements in the most general case. Next, we derive detailed results for non-orthogonal weak measurements and then for orthogonal weak measurements.

Frequently the system observable $\mathbf{A}$ acts on a finite-dimensional Hilbert space; it is convenient to redefine $\mathbf{A}$ and $g$ such that
$\mathbf{A}$ becomes a dimensionless operator with a unit norm (i.e.,
$\mathbf{A} \rightarrow \mathbf{A}/\|\mathbf{A}\|$), and $g \rightarrow g \|\mathbf{A}\|$. Throughout this article, we shall follow this convention.

\section{Weak-interaction condition}
It should be pointed out that AAV's original condition (\ref{AAVcondition})
is too strict as it may limit further applications in high-resolution experiments for signal amplification,
when the pre- and post-selected states are chosen to be nearly orthogonal.  In order to derive a formula for the regime
beyond (\ref{AAVcondition}), we shall refer to the condition
\begin{equation}\label{WM.}
g\Delta p \ll 1.
\end{equation}
as {\sl the weak-interaction condition}. This condition has no dependence on the arbitrary choices of the pre- and post-selected states, and we shall see later that it actually means that the interaction between the system and the measuring device is so weak that the state of the system stays almost unchanged during the interaction (however, the change of the device state is not restricted). For comparison, we refer to (\ref{AAVcondition}) as {\sl the original weak measurement condition}.
In the following we shall largely focus on the regime where the original weak measurement condition (\ref{AAVcondition})
is violated while the weak-interaction condition (\ref{WM.}) is still satisfied; this is the regime where the PPS are nearly or exactly orthogonal.
We shall first derive a general formalism with no restriction on the PPS.

\section{A general framework}

Now we set the stage for the general case.
We assume that the initial state (pre-selection) of the system is a general mixed state $\rho_s$ (instead of the pure state $\left| \psi _i \right\rangle$), and the initial state of the measuring device is also a general mixed state $\rho_d$.
Instead of a pure state $\left| \psi _f \right\rangle \left\langle \psi_f \right|$, we assume the post-selection is a general projection $\Pi_f$ onto a subspace.
The weak interaction (defined by the Hamiltonian $\mathcal{H}=g(t-t_0)\mathbf{A} \otimes p $) between the pre-selection and post-selection may not satisfy the original
weak measurement condition (\ref{AAVcondition}), but we assume it satisfies our weak-interaction condition (\ref{WM.}).
In the rest of this article, we shall derive the average shifts of the pointer position and momentum conditional on the general PPS in the above general case, and therefore understand what is really measured in such a generalized weak measurement scenario.

The weak value of an observable $\mathbf{A}$ can be naturally extended for the general pre-selected state $\rho_s$ and the general post-selection $\Pi_f$ as \cite{WM09}
\begin{equation}
\langle \mathbf{A} \rangle_w \equiv \frac{\text{tr} ( \Pi_f \mathbf{A} \rho_s )}{\text{tr} ( \Pi_f \rho_s )}
\end{equation}
which reduces to AAV's original definition of weak values (\ref{weak value}) when the PPS are pure states.  For the convenience of discussion, we also define the following notations
\begin{equation}
\langle \mathbf{A} \rangle_w^{m,l} \equiv \frac{\text{tr} (\Pi_f \mathbf{A}^m \rho_s \mathbf{A}^l)}{\text{tr} (\Pi_f \rho_s)} \label{generalizedWV}
\end{equation}
which can be viewed as generalized high-order weak values when $\text{tr} (\Pi_f \rho_s) \neq 0$.
One easily has $\langle \mathbf{A} \rangle _w^{1,0}=\langle \mathbf{A} \rangle_w$,
$\langle \mathbf{A} \rangle _w^{m,l}= (\langle \mathbf{A} \rangle _w^{l,m})^*$,
and $\langle \mathbf{A} \rangle _w^{n,n} \geq 0$.
When the PPS are pure states, it is obvious that
$\langle \mathbf{A} \rangle _w^{m,0} = \langle \mathbf{A}^m \rangle _w$,
$\langle \mathbf{A} \rangle _w^{m,l} = \langle \mathbf{A}^m \rangle _w (\langle \mathbf{A}^l \rangle _w)^*$, and $\langle \mathbf{A} \rangle _w^{n,n} = |\langle \mathbf{A}^n \rangle _w|^2$.

The time-evolution operator corresponding to the interaction Hamiltonian $\mathcal{H}=g \delta (t-t_0)\mathbf{A} \otimes p $ is given by $\mathcal{U}=e^{-ig\mathbf{A}p}$, where
the observable $\mathbf{A}$ we like to measure weakly is dimensionless and has unit norm, since we already redefined $g$ and $\mathbf{A}$.
The overall state after the interaction is given by
\begin{equation}
\rho' = \mathcal{U}\rho\mathcal{U^\dagger}
= e^{-ig\text{\bf ad} \mathbf{A}p}\circ \rho .
\end{equation}
Here $\text{\bf ad} \Omega \circ \Theta$ defines the adjoint action of $\Omega$ on $\Theta$ as ${\bf ad} \Omega\circ \Theta \equiv \Omega \Theta - \Theta \Omega=[\Omega, \Theta ]$, and similarly
${\bf ad}^n \Omega \circ \Theta \equiv {\bf ad} \Omega \circ ( {\bf ad}^{n-1} \Omega \circ \Theta ) = [\Omega, {\bf ad}^{n-1} \Omega \circ \Theta ]$.
We can expand $\rho'$ as
\begin{eqnarray}
\rho' &=& \rho + \sum^{+\infty}_{n=1}{\frac{(-ig)^n}{n!}\text{\bf ad}^n\mathbf{A}p\circ\rho} \label{adjoint operator expansion}  \\
&=& \rho_s \otimes \rho_d + \sum^{+\infty}_{n=1} \frac{(-ig)^n}{n!}
\sum^{n}_{k=0} (-1)^k \left(^n_k\right) \nonumber \\
& & \cdot (\mathbf{A} \otimes p)^{n-k} \rho_s \otimes \rho_d
(\mathbf{A} \otimes p)^{k}   \label{overallstateafterWI}
\end{eqnarray}
where $\left(^n_k\right)$ denotes the binomial coefficient.
The state of the system after the interaction is given by
\begin{equation}
\rho'_s = \text{tr}_d\rho'
=\rho_s + \sum^{+\infty}_{n=1}{\frac{(-ig)^n \langle p^n\rangle}{n!}\text{\bf ad}^n\!\! \mathbf{A} \circ \rho_s }
\end{equation}
with $ \langle p^n \rangle \equiv tr (  p^n \rho _d )$.
If $g\Delta p \ll 1$, or more generally
\begin{equation}\label{extended condition}
g\sup\{\sqrt[n]{\langle p^n\rangle}\ , n\in \mathbb{N}\} \ll 1 \quad
\end{equation}
the state of the system will stay almost unchanged, rather than collapsing after the interaction. This is the reason we call $g\Delta p \ll 1$ the weak-interaction condition.

The subsequent post-selection is generally represented by a projection $\Pi_f $ onto a subspace of the system's Hilbert space.
The density matrix of the device after the post-selection is given by
$\rho'_d =  \text{tr}_s(\Pi_f \rho' )/ \text{tr}(\Pi_f \rho') $, where $\text{tr}_s$ ($\text{tr}$) denotes the trace over the system (whole) Hilbert space.  From (\ref{overallstateafterWI}), we have
\begin{eqnarray}
\rho'_d &=& \frac{1}{\mathcal{N}} \{
\text{tr}_s ( \Pi_f \rho_s ) \cdot \rho_d + \sum^{+\infty}_{n=1} \frac{(-ig)^n}{n!}
\sum^{n}_{k=0} (-1)^k \left(^n_k\right) \nonumber \\
& & \cdot \text{tr}_s (\Pi_f \mathbf{A}^{n-k}  \rho_s \mathbf{A}^{k} ) \cdot p^{n-k} \rho_d
p^{k}  \} \label{rhodeviceafter}
\end{eqnarray}
where the normalization factor $\mathcal{N}$ gives the success probability of post-selection \begin{eqnarray}
\mathcal{N} &=& \text{tr}(\Pi_f \rho') = \text{tr}_s (\Pi_f \rho_s ) +
\sum^{+\infty}_{n=1}\frac{(-ig)^n}{n!}\langle p^n\rangle \nonumber \\
&\cdot & \sum^{n}_{k=0}(-1)^k \left(^n_k\right) \text{tr}_s (\Pi_f \mathbf{A}^{n-k} \rho_s \mathbf{A}^k ) .
\end{eqnarray}
The average shift of a pointer observable $\hat{o}$ is given by $\delta \hat{o} \equiv tr(\hat{o} \rho^{\prime}_d) -tr(\hat{o} \rho_d)$. We shall derive the average shifts of the pointer
position and momentum for two different cases: the non-orthogonal weak measurement and the orthogonal weak measurement.

\section{Non-orthogonal weak measurement}
Now we consider the case when the weak-interaction condition (\ref{WM.}) is satisfied and
$tr_s (\Pi_f \rho_s ) \ne 0$.  A weak measurement
in this regime will be referred to as {\sl a non-orthogonal weak measurement}.
The PPS we consider here do not have to be pure states. However, even if the PPS are pure states,
the non-orthogonal weak measurement scenario considered here extends AAV's original weak measurement. The non-orthogonal weak measurement
includes not only
AAV's original weak measurement as a special case, but also the case when the PPS are nearly (but not yet) orthogonal, i.e., the case when the original weak measurement condition (\ref{AAVcondition}) is usually violated due to small overlap: $\text{tr}_s (\Pi_f \rho_s )$.

It is convenient to write $\mathcal{N} =\text{tr}_s (\Pi_f \rho_s ) \cdot \mathcal{Z}$ with
\begin{equation}\label{Z}
\small
\mathcal{Z}=1+\sum^{+\infty}_{n=1}{\frac{(-ig)^n}{n!}\langle p^n \rangle \sum^{n}_{k=0}{(-1)^k \left(^n_k\right)} \langle \mathbf{A}\rangle_w ^{n-k,k}}
\end{equation}
where $\langle \mathbf{A}\rangle_w ^{n-k,k}$ is defined according to (\ref{generalizedWV}).
For pure PPS, $\langle \mathbf{A}\rangle_w ^{n-k,k} = \langle \mathbf{A}^{n-k}\rangle_w \langle \mathbf{A}^{k}\rangle_w^*$, and thus $\langle \mathbf{A}\rangle_w ^{1,1} = | \langle \mathbf{A}\rangle_w |^2$.
From (\ref{rhodeviceafter}), we obtain
\begin{equation}\label{outcomes}
\small
\rho'_d = \mathcal{Z}^{\!-1}\!\!\left[ \rho_d + \sum^{+\infty}_{n=1}\frac{(-ig)^n}{n!}\sum^{n}_{k=0}(-1)^k \left(^n_k\right)
\langle \mathbf{A}\rangle_w ^{n-k,k} p^{n-k} \rho_d p^k \right].
\end{equation}
With the weak-interaction condition \eqref{WM.}, we can neglect high-order terms of $g$,
\begin{eqnarray}
\rho'_d &\approx &
C \cdot \{ \rho_d - ig\left(\mathbf{A}_w p\rho_d - \mathbf{A}_w^*\rho_d p\right)  \nonumber \\
& + & g^2 ( \langle \mathbf{A}\rangle_w ^{1,1} p\rho_d p
-\frac{1}{2}{\langle \mathbf{A}^2\rangle}_w p^2 \rho_d -\frac{1}{2}{\langle \mathbf{A}^2\rangle}_w^*\rho_d p^2  )  \} \label{outcomes1} \\
C&=& [1 + 2g \langle p \rangle \Im \mathbf{A}_w + g^2 \langle p^2 \rangle \left( \langle \mathbf{A}\rangle_w ^{1,1} - \Re{\langle \mathbf{A}^2\rangle}_w  \right) ]^{-1}
\end{eqnarray}
The average shifts of $q$ and $p$ due to weak measurement are given by (to the second order of $g$)
\begin{eqnarray}
\delta q &=& tr(q \rho_d^{\prime}) - tr(q \rho_d) \nonumber \\
&=& C\{ g\Re \mathbf{A}_w + g \Im \mathbf{A}_w \langle \lbrace (q-\langle q\rangle), (p-\langle p\rangle) \rbrace \rangle \label{outcomes21} \\
& +&g^2(\langle pqp\rangle-\langle p^2\rangle\langle q\rangle) \left( \langle \mathbf{A}\rangle_w ^{1,1} - \Re\langle \mathbf{A}^2\rangle_w \right)+g^2 \langle p\rangle \Im\langle \mathbf{A}^2\rangle_w \}   \nonumber  \\
\delta p &=& tr(p \rho_d^{\prime}) - tr(p \rho_d) = C \{ 2g\Im \mathbf{A}_w \text{var}p   \nonumber  \\
&+& g^2 (\langle p^3\rangle-\langle p^2\rangle\langle p\rangle)
\left( \langle \mathbf{A}\rangle_w ^{1,1} - \Re\langle \mathbf{A}^2\rangle_w \right) \}.    \label{outcomes22}
\end{eqnarray}
Here $\text{var} \hat{o}= tr(\hat{o}^2 \rho_d) - (tr(\hat{o}\rho_d))^2 $ and
$\text{var}^{\prime} \hat{o}= tr(\hat{o}^2 \rho_d^{\prime}) - (tr(\hat{o}\rho_d^{\prime}))^2 $.
In order to simplify the formulae, we further assume that the initial pointer state $\rho_d$ satisfies $\langle p \rangle =0$ and $\langle q \rangle =0$ (they are clearly satisfied by a pure initial pointer state that is an even wave function). Therefore we have
\begin{eqnarray}
\delta q &=& \frac{g\Re \mathbf{A}_w + g \Im \mathbf{A}_w \langle \lbrace q, p \rbrace \rangle}{1+g^2\text{var}p (\langle \mathbf{A}\rangle_w ^{1,1} - \Re{\langle \mathbf{A}^2\rangle}_w)}  \label{outcomes31}  \\
\delta p &=& \frac{2g\Im \mathbf{A}_w \text{var}p}{1+g^2\text{var}p (\langle \mathbf{A}\rangle_w ^{1,1} - \Re{\langle \mathbf{A}^2\rangle}_w)}
\label{outcomes32}
\end{eqnarray}

When $tr_s (\rho_s \Pi_f) \rightarrow 0$ (but $\neq 0$),
the average shifts of $q$ and $p$ may increase first, and
then decrease rapidly when $\langle \mathbf{A}\rangle_w ^{1,1} - \Re{\langle \mathbf{A}^2\rangle}_w$ becomes comparable to $1/{g^2\text{var}p}$. Therefore, for fixed $g$ and $\text{var} p$,
there may exist a maximum shift of a pointer quantity and an optimal overlap  $tr_s (\Pi_f \rho_s)$ to achieve the maximum shift.
We cannot increase the shift of the pointer quantity arbitrarily by varying the overlap between the pre- and post-selected states.
This is a very surprising result, and we shall discuss it below in more detail for the Stern-Gerlach experiment setup by AAV.

\begin{figure}
\centering
\includegraphics[width=6.5cm]{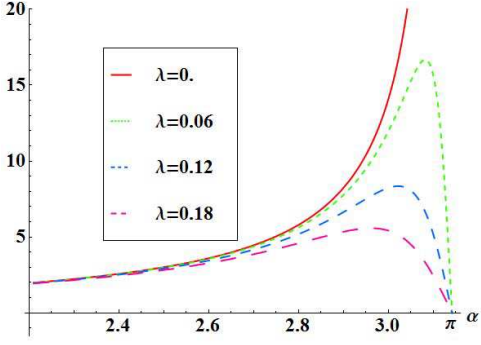}
\caption{(Color online) Outcome for a weak measurement of $\sigma _z$ under different interaction strength $\lambda$.}
\label{modify}
\end{figure}

Spin-1/2 particles with a pre-selected spin pointed in the direction \textbf{$\xi$} (which is on the xz plane)
travel along the y axis through an inhomogeneous (in the z direction) weak magnetic field, and then post-selected by a strong inhomogeneous
magnetic field in the $x$ direction  (See FIG. 1 in \cite{AAV}).  The Hamiltonian describing the weak interaction can be written as
$\mathcal{H}=g \delta(t-t_0)\sigma_z  \otimes z$, where $g=-\mu \frac{\partial B_z}{\partial z}$ depends on the weak magnetic field.
With the substitution $\mathbf{A} \rightarrow \sigma_z$, $p \rightarrow z$ and $q \rightarrow - p_z$ in (\ref{outcomes31}),
the average shift of the z component of the particle momentum $\delta p_z$ conditional on pre- and post-selection is obtained directly, which
in turn gives the measured value of the spin component $\sigma_z$:
\begin{equation}
\frac{\sin\alpha}{(1-\frac{\lambda^2}{2})\cos\alpha+1}  \label{SG}
\end{equation}
where $\alpha$ denotes the angle between \textbf{$\xi$} and $x$, $\lambda =|g|/ \Delta p_z =| \mu \frac{\partial B_z}{\partial z}| / \Delta p_z$
denotes the interaction strength, and $\Delta p_z =\sqrt{var p_z}$ denotes the standard deviation of $p_z$ in the initial (pre-selected) state, which
is related to the width $w=\Delta z$ for a Gaussian beam by $\Delta p_z =\frac{1}{2 w}$.
The weak-interaction condition (\ref{WM.}) requires $\lambda  \ll 1$.
The predicted weak measurement outcome of $\sigma_z$ as a function of $\alpha$ is
plotted in FIG. \ref{modify} for different values of $\lambda$.
If we let $\varepsilon = \pi - \alpha$, it follows that $\varepsilon$ is very small
when the PPS are close to being orthogonal.
AAV's original weak measurement condition (\ref{AAVcondition}) requires
$\lambda \ll \varepsilon$, which is quickly violated as $\varepsilon \rightarrow 0$.
Eq. (\ref{SG}) reduces to $\frac{2 \varepsilon}{\lambda^2 +\varepsilon^2}$ with $\varepsilon \rightarrow 0$.
One can also see that for a fixed value of $\lambda$, the maximum outcome is approximately (of the order of) $1/\lambda$, which is achieved when $\alpha \approx \pi - \lambda$.
Therefore, for a fixed $\lambda$, the measured value of the spin component may not be increased arbitrarily by decreasing the overlap between the PPS.
The AAV's original formula corresponds to the limiting case $\lambda \rightarrow 0$.

When the PPS are pure states and the overlap $|\langle \psi_f | \psi_i \rangle|$ is not small (compared with $g \Delta p$) so the original weak measurement condition (\ref{AAVcondition}) is satisfied, we have
$g^2\text{var}p (|\mathbf{A}_w|^2 - \Re{\langle \mathbf{A}^2\rangle}_w) \ll 1$; therefore, Eqs. (\ref{outcomes31}) and (\ref{outcomes32}) directly reduce to Eqs. (\ref{deltaq}) and (\ref{deltap}). AAV's original weak measurement is indeed a special case of the general scenario considered here.

\section{Orthogonal weak measurement}
Now we consider another limiting case of our non-orthogonal weak measurement, i.e., the case when
the pre-selection and the post-selection are orthogonal [i.e., $tr_s (\Pi_f \rho_s) = 0$] while
the weak-interaction condition Eq. \eqref{WM.} is still satisfied.
A weak measurement in this scenario shall be referred to as {\sl orthogonal weak measurement}. The original weak measurement formalism totally fails (even for pure PPS), and the weak values are even not defined.

Due to the weak interaction between the device and the system, the system state $\rho_s^{\prime}$ after the interaction could differ slightly from the pre-selected state $\rho_s$; therefore, the follow-up post-selection could have a nonzero success probability in general.
However, when either the pre-selected state $\rho_s$ or the post-selection $\Pi_f$ commutes with the observable $\mathbf{A}$, the success probability for post-selection is strictly zero.  Therefore, in the following discussion of orthogonal weak measurements, we assume that neither $\rho_s$ nor $\Pi_f$ commutes with $\mathbf{A}$.

Since for the orthogonal case $tr_s (\Pi_f \rho_s) = 0$, one immediately has $\Pi_f \rho_s = \rho_s \Pi_f =0$. Therefore, from (\ref{rhodeviceafter}), one has
\begin{eqnarray}
\rho'_d &=& \frac{g^2}{\mathcal{N}} \{
\text{tr}_s ( \Pi_f \mathbf{A} \rho_s \mathbf{A})\cdot p \rho_d p + \sum^{+\infty}_{n=1} \frac{(-ig)^n}{n!}
\sum^{n}_{k=0} (-1)^k \left(^n_k\right) \nonumber \\
& & \cdot \frac{\text{tr}_s (\Pi_f \mathbf{A}^{n+1-k}  \rho_s \mathbf{A}^{k+1} )}{(n+1-k)(k+1)} \cdot p^{n+1-k} \rho_d
p^{k+1}  \} \label{rhodeviceafterorthogonal}
\end{eqnarray}
where
\begin{eqnarray}
\mathcal{N} &=&
g^2 \{
\text{tr}_s ( \Pi_f \mathbf{A} \rho_s \mathbf{A}) \langle p^2 \rangle + \sum^{+\infty}_{n=1} \frac{(-ig)^n}{n!}
\sum^{n}_{k=0} (-1)^k \left(^n_k\right) \nonumber \\
& & \cdot \frac{\text{tr}_s (\Pi_f \mathbf{A}^{n+1-k}  \rho_s \mathbf{A}^{k+1} )}{(n+1-k)(k+1)}
\langle p^{n+2} \rangle\} .
\end{eqnarray}

Without loss of generality, we assume that the $g^2$ term in (\ref{rhodeviceafterorthogonal}) does not vanish, namely, $tr_s (\Pi_f \mathbf{A} \rho_s \mathbf{A}) \neq 0$ (similar discussion follows directly if the $g^n$ term is the lowest-order term).
(\ref{rhodeviceafterorthogonal}) can be re-written as
\begin{eqnarray}
\rho'_d &=& \mathcal{Z}_o^{-1}\langle p^2\rangle^{-1}\cdot   p \left[ \rho_d  + \sum^{+\infty}_{n=1}\frac{(-ig)^n}{n!} \cdot \right. \nonumber \\
&&\left. \sum^{n}_{k=0}(-1)^k \left(^n_k\right)
\langle \mathbf{A} \rangle_{ow}^{n-k,k}  p^{n-k} \rho_d p^k \right]p
\label{orthogonal outcomes}
\end{eqnarray}
with
\begin{equation}
\small
\mathcal{Z}_o = 1 + \sum^{+\infty}_{n=1}{\frac{(-ig)^n}{n!} \frac{\langle p^{n+2}\rangle}{\langle p^2 \rangle} \sum^n_{k=0}(-1)^k \left(^n_k\right) \langle \mathbf{A} \rangle_{ow}^{n-k,k}}
\end{equation}
where
\begin{equation}
\langle \mathbf{A} \rangle_{ow}^{m,l} \equiv \frac{\text{tr}_s (\Pi_f \mathbf{A}^{m+1}  \rho_s \mathbf{A}^{l+1} )}{(m+1) (l+1) \text{tr}_s (\Pi_f \mathbf{A}  \rho_s \mathbf{A} ) }
\end{equation}
can be viewed as the generalized weak values corresponding to (\ref{generalizedWV}) for the orthogonal case.

For simplicity, in the following we assume that the PPS are pure states, i.e.,
$\rho_s = |\psi_i \rangle \langle \psi_i |$ and $\Pi_f = |\psi_f \rangle \langle \psi_f |$;
and we further assume that
the initial wave function of the device is \textit{even} (or $tr (\rho_d p)=0$). One obtains
\begin{eqnarray}
\delta q &=& g \Re \mathbf{A}_{ow} + g \Im \mathbf{A}_{ow} \langle p \lbrace q,p \rbrace p \rangle / \langle p^2 \rangle \\
\delta p &=& 2g \Im \mathbf{A}_{ow} \langle p^4 \rangle / \langle p^2 \rangle  \\
\text{var}' q &=& \langle pq^2p \rangle /\langle p^2 \rangle \\
\text{var}' p &=& \langle p^4 \rangle / \langle p^2 \rangle
\end{eqnarray}
where $\langle \mathbf{A}\rangle_{ow} \equiv  \langle \mathbf{A}\rangle_{ow}^{1,0}=\frac{\langle \psi_f|\mathbf{A}^2|\psi_i \rangle}{2\langle \psi_f|\mathbf{A}|\psi_i \rangle}$ when the PPS are pure states.
Here, $\left\langle \hat{o} \right\rangle = tr (\hat{o} \rho_d)$ is the expectation value of $\hat{o}$ with respect to the initial state.

\begin{figure}
\centering
\includegraphics[width=7.5cm]{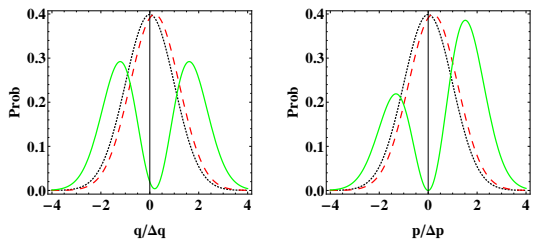}
\caption{(Color online) Probability density as a function of $q/\Delta q$ (left) and $p/ \Delta p$ (right).
The dotted black lines represent the initial probability density function of the device which are assumed to be Gaussian,
the solid green lines represent the probability density function in an orthogonal weak measurement with $\mathbf{A}_{ow}=0.2+0.1i $, and the dashed red lines represent those in a non-orthogonal weak measurement with  $\mathbf{A}_{w}=0.2+0.1i$ for comparison. }
\label{distribution}
\end{figure}

When the initial pointer state is a Gaussian, the results become even simpler
\begin{eqnarray}
\delta q = g \Re \mathbf{A}_{ow}  \quad \quad
\delta p = 6g \Im \mathbf{A}_{ow} \text{var}p  \\
\text{var}' q = 3 \text{var}q   \quad \quad
\text{var}' p = 3 \text{var}p
\end{eqnarray}
where $\text{var}' q$ and $\text{var}' p$ are the variances of $q$ and $p$ in the final pointer state after the post-selection. The final probability distribution function
has two peaks at $ p_{\max} \approx g\Im \mathbf{A}_{ow}(\Delta p)^2 \pm\sqrt{2}\Delta p$ in the $p$ space,
and at $q_{\max} \approx g\Re \mathbf{A}_{ow}\pm \sqrt{2}\Delta q $ in the $q$ space (see Fig.~\ref{distribution}).
The variances of $q$ and $p$ are generally enlarged because the wave function spreads out with double peaks after the orthogonal weak measurement.
This feature is exactly confirmed by Ritchie \textit{et al.}'s experiment (see FIG. 2c in \cite{Ritchie}), with two almost symmetrical peaks observed, since the orthogonal weak value vanishes in their case.
The double hump shown in Fig.~\ref{distribution} also looks similar to Fig. 4 in \cite{Duck89} and Fig. 2 in \cite{SDWJH}; this similarity is expected because the regimes of interest coincide.

\section{Concluding remarks} 
In summary, we have extended the weak measurement theory beyond the original formalism, presented a complete treatment and derived results for both non-orthogonal and orthogonal weak measurements. Our results for non-orthogonal weak measurements work well in the regime where the previous formalism of weak measurements starts to fail.  We found that the measurement outcomes cannot increase arbitrarily by decreasing the overlap between the pre- and post-selected states. We have also obtained interesting results for measurement outcomes in orthogonal weak measurements, a case when the original weak measurement formalism completely fails and the weak values are even not defined. The interesting quantities we found in orthogonal weak measurements play a role surprisingly similar to that of the weak values in the original weak measurement context.
The general results we have derived here not only extend the idea of weak measurement to the general case, but also provide a framework for experimental guidance and further applications to signal amplification.

It should be pointed out that there are several recent works that seem related. Contextual values \cite{DAJ} and modular values \cite{KV} of an observable
are introduced to generalize the usual concept of weak values to an interaction of any strength, while
we deal with a weak interaction directly and present a straightforward approach to derive the measurement outcomes for general pre- and post-selection in real experiments.
For a Sagnac interferometer \cite{SDWJH} with continuous phase amplification and
the original Stern-Gerlach setup \cite{Geszti}, specific results are obtained in the regime when the PPS are nearly orthogonal. However,
our general results do not refer to a specific experimental setup; they apply to any experimental setup and to any PPS, even when the PPS are exactly orthogonal. Most of our results also hold for
the case when the initial state is a general mixed state $\rho_s\otimes \rho_d$ and the post-selection is a projection onto a subspace $\Pi_f$ of the system.

\begin{acknowledgements}
The authors wish to acknowledge support from the NNSF of China (Grant No. 11075148), the CUSF, the CAS, and the National Fundamental Research Program.
\end{acknowledgements}

\end{document}